# Self-organized Voids Revisited: Experimental Verification of the Formation Mechanism[*]

Song Juan(宋娟)[a†], Ye Jun-yi(叶俊毅)[b], Qian Meng-di(钱梦迪)[b], Luo Fang-fang(骆芳芳)[c], Lin Xian (林贤)[b], Bian Hua-dong(卞华栋)[b], Dai Ye(戴晔)[b], Ma Guo-hong(马国宏)[b], Chen Qing-xi (陈庆希)[c], Jiang Yan(姜燕)[a], Zhao Quan-zhong(赵全忠)[c], Qiu Jian-rong(邱建荣)[d]

[a]School of Material Science and Engineering, Jiangsu University, Zhenjiang 212013, China

[b]Physics Department, Shanghai University, Shanghai 200444, China

[c]State Key Laboratory of High Field Laser Physics, Shanghai Institute of Optics and Fine Mechanics, Chinese Academy of Sciences, Shanghai 201800, China

[d]Key Laboratory of Specially Functional Materials of Ministry of Education and Institute of Optical Communication Materials, South China University of Technology, Guangzhou, Guangdong 510640, China

E-mail: ddvsh@163.com

**Abstract** In this paper, several experiments were conducted to further clarify the formation mechanism of self organized void array induced by a single laser beam, including energy-related experiments, refractive-index-contrast-related experiments, depth-related experiments and effective-numerical-aperture experiment. These experiments indicate that the interface spherical aberration is indeed responsible for the formation of void arrays.

## 1. Introduction

Due to its powerful three-dimensional region-selectable micromachining ability, femtosecond laser has been widely applied in preparation of micro-/nano-structures. Much research progress has been achieved in this field in the past decades, such as polarization-dependent nanoripples,[1-3] optically produced arrays of planar nanostructures,[4] nanostructures inscribed by non-reciprocal ultrafast laser writing,[5] axial dot arrays imprinted by a single pulse[6-7] and so on. These novel micro-/nano-structures have been employed to develop new micro-optics devices including radially polarized optical vortex converters,[8] near-infrared photoelectric detectors,[9-10] optical vortex generators,[11] etc. Moreover, self-assembled polarization-dependent nanogratings induced inside porous glass have recently been reported to be reduced to a single sub-50nm-wide nanochannel by the near-threshold-ablation technique, and this kind of single nanochannel has further been employed as building blocks of a 3D micro-nanofluidic device which has been used to demonstrate DNA analysis.[12-13] These research progress opens new opportunities of self-assembled nanostructures in 3D micro-nanofluidic application. Precision manufacturing of functional micro-devices by assembling various kinds of microstructures is based on the prerequisite that the microstructures can be flexibly controlled and adjusted as desired, and clarification of the formation mechanism is the key to refining the controlling factor. As an example, self-organized void array, first reported by S. Kanehira,[14] are expected to be used as photonic crystals,[15] diffraction devices [6] and optical storage units,[7] but the structure paramenter (e.g. void-void spacing, void diameter uniformity, and total length) of the void array still cannot be accurately changed because of the lack of generally-accepted formation mechanism. The void array is usually induced by tightly focusing ultrashort pulses from the ambient environment (AE) into the sample. In this light-propagation issue, the interface spherical aberration (ISA) and the nonlinear effects (NE) both have to be taken into account. The ISA which is caused by the AE/sample refractive index mismatch can destroy the concentricity of the focused incident light and lead to multiple focuses in the sample. The NE includes kerr self-focusing, electron plasma defocusing, multiple-photon absorption as well as plasma absorption, and the balance and competition between them can also result in multiple focuses. Although we proposed that the AE/sample ISA but not NE led to the self-assembly action of the voids,[16] nothing but a simulated fluence has been presented to support this viewpoint. In this paper, we attempt to find some experimental evidences to figure out whether the ISA or NE is the main reason for void array formation. As known to all, the NE is dependent on the laser peak power, and the ISA amount is determined by the refractive index contrast between AE and the sample, the focal depth below the sample surface and the effective numerical aperture (NA) of the objective lens. Four-group experiments related to the above factors were conducted to further make clear the formation mechanism.

---

[*] Project supported by National Basic Research Program of China (Grant Nos. 61205128 and1102075) and the Research Foundation for Advanced Talents of Jiangsu University (No. 09JDG022).
[†] Corresponding author. E-mail: ddvsh@163.com

2. **Experiments**

In our experiments, a pulse train containing pulses with 800 nm wavelength, 120 fs duration and 1 kHz repetition rate, launched from a Ti: sapphire regenerative amplifier laser system ( Spitfire，Spectra- Physics Co.) , was focused by an objective lens into the bulk of the glass sample mounted on a three-dimensional motorized stage. The pulse energy and pulse number were controlled respectively by a neutral filter and an electronic shutter. The sample was a piece of fused silica glass with four facet polished. In our study, a dry objective lens with NA=0.9 and an oil-immersed lens with NA=1.45 were both employed for comparison. An adjustable aperture was used for truncating the beam to obtain different effective NA of the lens. The details of our experiments will be introduced in the following subsections. Besides, since the void-array structures are once reported to be fabricated by even a single pulse and the period property of the void array (e.g. void-void interval) are influenced not so much and not so regularly by the pulse number,[6,17] the pulse number effect is not an key issue for investigating the periodicity origin of the void array and will be not addressed in the following experiments and discussions.

**2.1 Energy-controlled experiments**

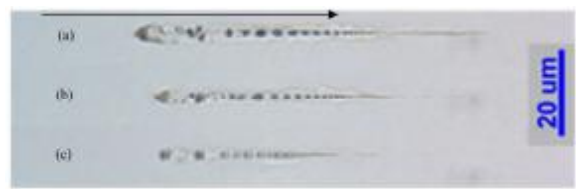

Fig.1 Void arrays induced at the focal depth of 100 μm with different pulse energy of

15 μJ (a), 10 μJ (b) and 5 μJ (c). The arrow indicates the laser propagation direction.

As femtosecond laser micromachining in the bulk of material involves nonlinear propagation issues of laser pulses, it is tempting to ascribe the self-organization action of the voids to a succession of the kerr self-focusing and plasma defocusing phenomena. The self-focusing phenomena strongly depends on the peak power of the pulse, so the pulse energy of 15 μJ, 10 μJ and 5 μJ, which has the peak power of 117 MW, 78 MW and 39 MW severely exceeding the self-focusing critical Power in fused silica (2.7 MW), were chosen to investigate the influence of the nonlinear effects. The generated void arrays are displayed in Fig. 1, where a train of 63 laser pulses was focused 100 μm beneath the sample surface through a 100×/NA=0.9 dry objective lens. The image shows that the laser beam with larger single-pulse energy inscribed a longer structure with more void number.

**2.2 Refractive-index-contrast experiment**

It is known that the ISA happens when light passes through an interface of two media with different refractive indices. P. Török theoretically analyzed the diffraction of electromagnetic waves for light focused by a high-NA lens from the first material into the second material.[18] In their theory, the amount of ISA is expressed by aberration function $\Psi = h(n_2 \cos\theta_2 - n_1 \cos\theta_1)$, which indicate that the refractive index difference of two media $n_1 / n_2$, the focal depth **h** and the convergence angle θ of the focused light have a combined effect on the ISA. In this subsection, the amount of refractive index difference was first considered. A NA=0.9 dry objective lens designed for working in air and a NA=1.45 oil-immersion objective lens required to work in oil were respectively used for comparison. For the same fused silica sample, two different lenses used correspond to two different interface cases, namely, air-silica interface with 1/1.46 refractive index contrast and oil-silica interface with 1.515/1.46 contrast. Fig. 2 shows the microstructures ablated in the two cases, where the pulse energy of 5 μJ, the irradiation pulse number of 63 and the focusing depth of 100μm are set as the same. It is apparent that the air-silica case induced a much longer void array than the oil-silica case.

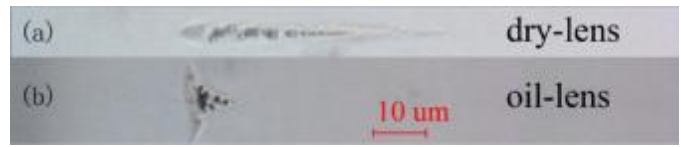

Fig. 2 Structures induced respectively by focusing 63 pulses at a depth of 100 μm inside fused silica through a dry lens with NA=0.9 (a) and an oil-immersion lens with NA=1.45 (b). Pulse energy was 5 μJ in both cases.

## 2.3 Depth-related experiments

As listed above, the ISA is supposed to be also dependent on the focal depth in the sample. For both the dry-lens case and the oil-lens case, depth was adjusted to investigate whether structures were changed correspondingly. In consideration of the significantly different working distance of two lenses (900 μm for dry lens and 170 μm for oil lens), focal depths of 100 μm, 200 μm and 300 μm were chosen for dry lens, while focal depths of 20 μm, 60 μm and 100 μm were set for oil lens. Irradiation pulse number was kept as the same to be 63. Pulse energy was respectively fixed at 15μJ and 0.5μJ for dry lens and oil lens. The obtained structures are presented in Fig. 3. It is clear that the scale of the void array increased significantly in terms of void array length and void number for the dry-lens case, but the structures changed little for the oil-lens case.

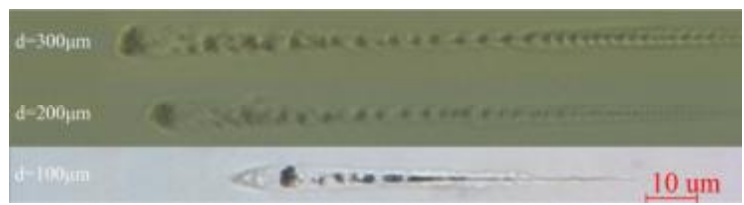

(a)

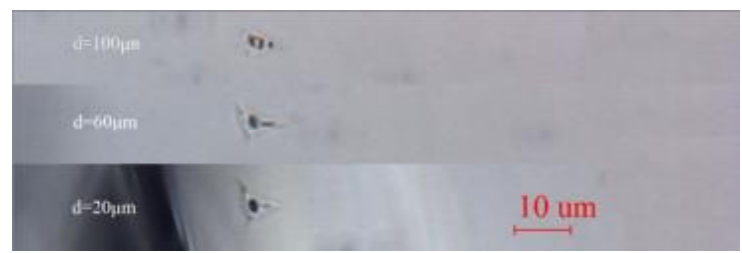

(b)

Fig. 3 Effect of the focal depth on the structure: (a) the void arrays generated by focusing laser pulses of 15 μJ at three depths of 300 μm, 200 μm and 100 μm. (b) the void arrays inscribed by focusing laser pulses of 0.5 μJ at depths of 100 μm, 60 μm and 20 μm. Number of pulse for irradiation in all subfigures are 63.

## 2.4 Tunable effective-NA experiments

Back to the aberration function again, the convergence angle of light has innegligible impact on the ISA. Theoretically, when the aperture of the objective lens is completely filled with a light wave, the half angle of light cone coming out from lens is determined by the nominal NA through $sin^{-1}(NA/n_{medium})$. In our experiments, an adjustable aperture with its center aligned with the optical axis, was inserted in the light path to truncate the incident beam to change the effective beam diameter incident on the lens. In this case, half angle of light cone is calculated by effective NA rather than nominal NA. The original beam diameter is 4mm, and the aperture was set as 0.8 mm, 1.6 mm and 2 mm respectively for experiments. With the pulse energy of 15 μJ and the irradiation pulse number of 63 kept fixed, the void arrays generated by focusing aperture-truncated laser beam 300μm beneath the sample surface are displayed in Fig. 4. It is clear that a larger aperture, namely a higher effective NA, makes a longer void array and a larger void number.

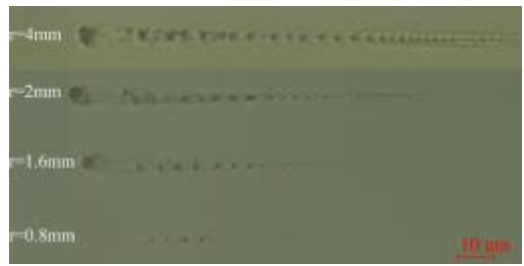

Fig. 4 Void arrays induced at a focal depth of 300μm by the laser beam truncated with apertures of 4 mm, 2 mm, 1.6 mm, 0.8 mm. In all subfigures, the pulse energy of 15 μJ and pulse number of 63 is used.

## 3. Discussions

It is generally accepted that onset of self-focusing is dependent on the laser power rather than the intensity. With the laser peak power exceeding self-focusing critical power, the self-focusing caused by kerr effect and the defocusing caused by plasma occurs simultaneously in fused silica. Although the structures in Fig. 1 show some correlations with laser peak power, it is insufficient to draw the conclusion that nonlinear effects result in the self-organization action. A physical model considering both the ISA theory and the NE of pulse propagation, was once proposed by us and is also adopted in this paper.[16] Moreover, since the femtosecond laser pulse inherently contains a wide spectrum ranging from about 790 nm to 810 nm, group velocity dispersion(GVD) is also incorporated in our model. In addition to GVD effects, the wide spectrum characteristics also may cause interface chromatic aberration. However, the relative refractive index change in the whole spectrum is both less than $5 \times 10^{-4}$ for either air or fused silica, chromatic aberration can be ignored compared to the spherical aberration. Hence, in all of the following analyses, an average refractive index in the 790nm-810nm wavelength range is used for the interface spherical aberration. According to the above model, the on-axis fluence distributions were obtained for pulse energy of 5 μJ and 10 μJ, as presented in the left panel of Fig. 5. Obviously, the periodically modulated fluence along the laser propagation axis is roughly consistent with the structure features of the void arrays. The fluence recalculated through the model with only nonlinear effects considered are also displayed in the right panel of Fig. 5 for comparison.[16, 19] It may be that the combined effects of defocusing caused by the natural diffraction and plasma surpassed the kerr self-focusing effect, so there are no short-period multiple focuses which are expected to result from a succession of kerr self-focusing and plasma defocusing phenomena. The comparisons between two models clearly indicate that the fluence periodicity is most probably caused by ISA rather than NE. Besides, Fig. 5(a) and Fig. 5(b) show that the fluence at pulse energy of 10 μJ are larger than the fluence at pulse energy of 5 μJ. Taking into account the inherent microexplosion threshold for fused silica, simulated results predict that more microexplosions will happen for higher energy, which is in agreement with the experimental results of Fig.

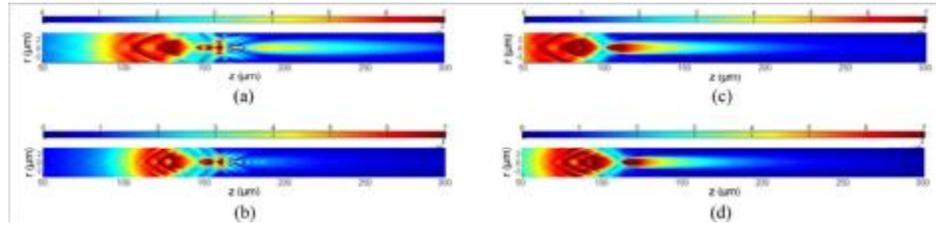

Fig. 5 Simulated fluence distribution for pulse energy of 10μJ (a) (c) and of 5μJ (b) (d). Both nonlinear effects and interface SA are considered in subfigures (a) and (b). Only nonlinear effects are taken into account in (c) and (d).

To further verify the formation mechanism to be ISA, experiments specially designed for ISA in section 2.2, 2.3 and 2.4 are paid more attentions. From the view of geometrical optics, the ISA can be interpreted as the multiple focuses caused by different convergence degrees of peripheral rays and paraxial rays when the concentric convergent rays passing through an interface, as shown in Fig. 6. The focus number and range of focal region is totally determined by the refractive index contrast between $n_1$ and $n_2$, the focal depth $h = |OA|$ below the interface and the half-angle $\theta_{max}$ of the maximum cone of light striking on the interface. Theoretically, increasing either of $|n_2-n_1|/n_1$, $|OA|$, and $\theta_{max}$ with other two kept fixed will obtain a longer focal region and focuses.

For experiments in section 2.2, the dry lens with NA=0.9 and the oil lens with NA=1.45 was used respectively for focusing the laser beam at the same depth. In consideration of the immersion media in two cases (air $n=1$ and oil $n=1.45$ respectively), half angle $\theta_{max}$ of maximum light cone is calculated to be 1.20 rad and 1.25 rad, which can be approximately seen as the same. The only notable difference is the refractive index contrast of $|1.46-1|/1$ and $|1.46-1.52|/1.46$ for two cases. The longer void array in dry lens is most probably ascribed to the higher refractive index contrast, which is in agreement with the theoretical prediction.

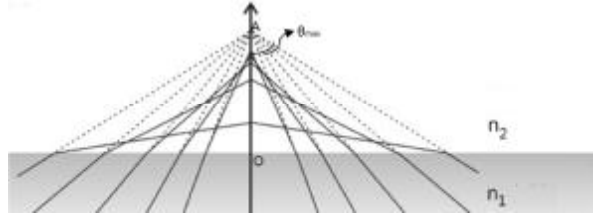

Fig. 6 Geometrical interpretation of interface SA

The rigorous simulation results based on P. Török's ISA theory are shown in Fig. 7. It is intuitive that for the dry-lens case, the focal region is much longer and the electrical field strength difference between multiple periodically modulated peaks is far smaller. These simulation shows that a long void array structure will be expected in dry-lens experiments. These provide a basic explanation for the difference between the dry-lens and the oil-lens shown in Fig. 2. It is worth noting that the periodic fluence distribution in oil-lens case seems to be opposite to that in dry-lens case in terms of the monotonic change direction of the fluence peak. It is possibly because that the slightly-aberrated but tightly-focused light ray in the oil-lens case makes NE greatly adjust the fluence distribution modulated periodically by ISA. Specifically speaking, on the one hand, as higher fluence peak will generate more free electrons and then be absorbed more by plasmas in contrast to the lower fluence peak, the highest fluence peak may be converted to lowest peak; on the other hand, the kerr self-focusing and the plasma defocusing may lead to the rearrangement of the fluence peak. These two reasons join up to lead that the resulting monotonic change direction of the fluence peak is inversed.

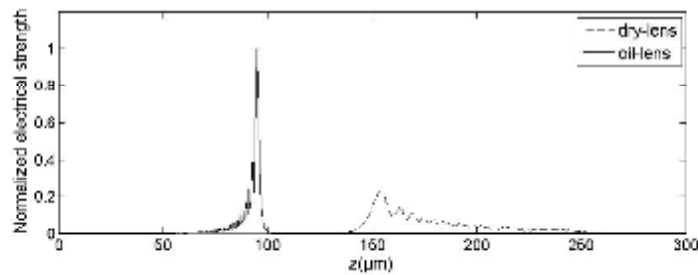

Fig. 7 Simulated axial electrical field strength for the dry-lens case and the oil-lens case. In both cases, depth of 100μm was adopted for the simulation based on ISA theory.

The effect of different focal depths in Fig. 3 can also be qualitatively understood in the context of ISA, since a larger focal depth naturally results in a wider focal region as shown in Fig. 6. The quantitative simulations for the impact of focal depths in dry-lens case are also given in Fig. 8 (a), where for a deeper depth, the electric strength distribution spreads over a wider axial range and the differences between adjacent peaks become relatively smaller. To be compared, the simulated results for oil lens are also displayed in Fig. 8 (b). It is obvious that due to the far smaller refractive index contrast in oil-lens case, the amount of ISA is much smaller so that the multi-focus phenomenon depends less on the focal depth. Therefore, the experimental results can be well predicted by the simulation results of ISA theory.

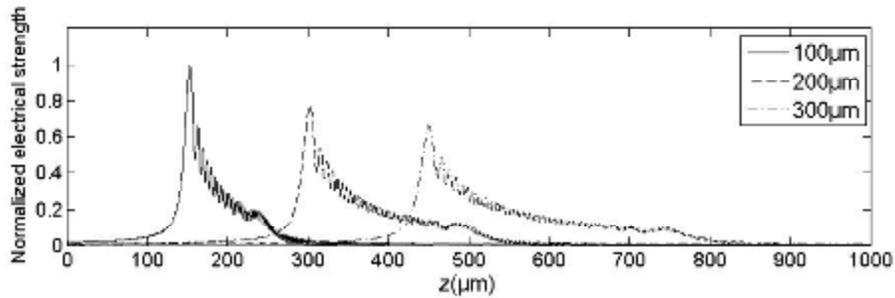

(a)

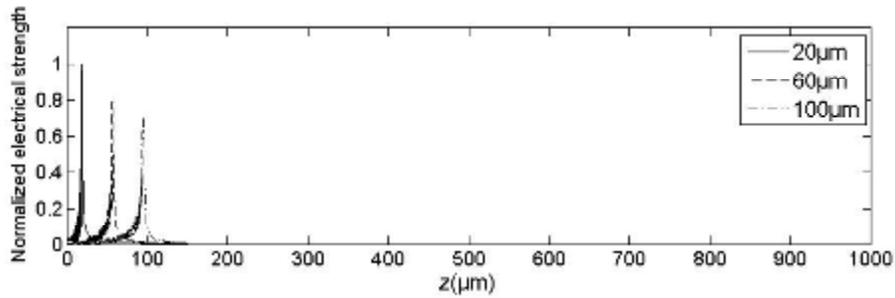

(b)

Fig. 8 Simulation results for investigating the impact of focal depth on the axial electrical field strength. (a) dry-lens case (b) oil-lens case

ISA means that peripheral rays and paraxial rays contributes to different axial focuses. Blocking of peripheral rays is supposed to reduce the focus number and focal region length. In section 2.4, controlling the amount of the blocked rays through an adjustable aperture was expected to adjust the void number as well as void array length, and the experimental results in Fig. 4 did verify that. In Fig. 9, the influence of the adjustable aperture on the light field distribution was quantified by P. Török's SA theory, by taking into account that the aperture diameter of 2 mm, 1.6 mm and 0.8 mm corresponds to the half angle of light cone of 0.838 rad, 0.726 rad and 0.4182 rad. Clearly, the simulated electrical-field-strength distributions are roughly consistent with the experiment results. However, truncating the beam with an adjustable aperture may arouse the doubt that maybe it is power reduction rather than NA shrinking that lead to the changing tendency of void arrays. Hence, with iris aperture fixed as 0.8 mm, laser pulse energy striking at the iris was increased from 15 μJ to 45 μJ to observe the effect of effective pulse energy on the void array. The results are displayed in Fig. 10. Compared with the void array induced by laser pulse energy of 15 μJ with the adjustable aperture removed, with effective NA kept as a low constant value of *sin(arctan(0.8/0.9))* , the structures change little with the increase of pulse energy and are no more than few voids around the top section. For the case of pulse energy of 45 μJ where the effective pulse energy after the adjustable aperture is 9.6 μJ (Fig.10 (h)), its void array is roughly equivalent in length to the void array produced by laser pulses of 5 μJ (Fig.1 (c)). From experiments in Fig. 1 and Fig 3, increase of pulse energy and focal depth among the parameter range involved by us will definitely elongate the void array, so the void array induced by the pulse energy of 9.6 μJ and the focal depth of 300 μm in Fig .10 (h) should have being longer than the void array fabricated by the pulse energy of 5 μJ and the focal depth of 100 μm in Fig. 1 (c). Therefore, the nearly equivalent length of both cases is unexpected and may only result from smaller NA in Fig. 10(h). These experimental facts convince us that effective NA has a significant effect on the void arrays, which is just the conclusion of ISA.

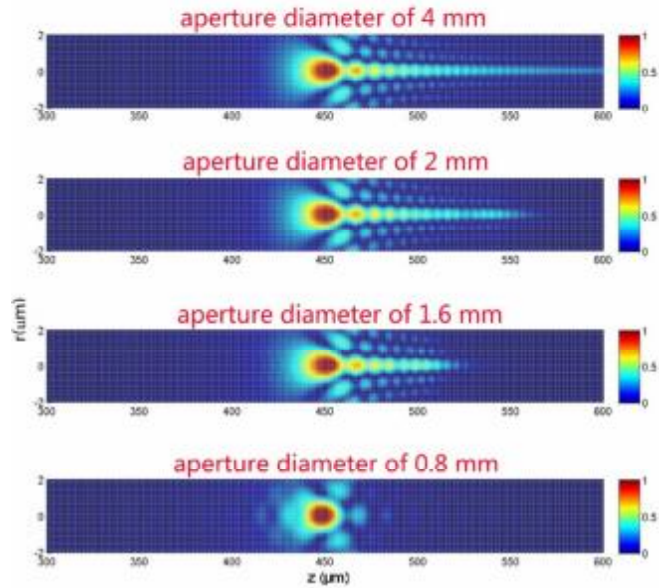

Fig. 9 Simulation results of the effect of the adjustable aperture diameter on the electrical field strength. From the top down, apertures of 4 mm, 2 mm, 1.6 mm and 0.8 mm are set for simulation. The focal depths used in simulation are all 300 μm.

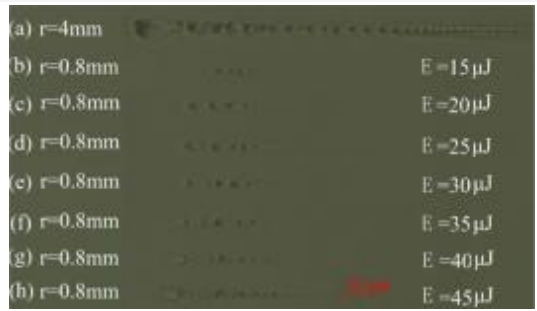

Fig.10 Influence of different pulse energy on the void arrays induced by a 0.8 mm-aperture-truncated laser beam. The focus depth is 300 μm for all subfigures.

## 4. Conclusion

In this paper, the influence on the void array of the laser pulse energy, the refractive index contrast between the lens-immersion medium and the sample, the focal depth and the effective NA was in detail investigated. Energy-related experiments and energy-related simulation indicate that NE plays a little role in the self-organization action of voids. Experiments specially designed for testing ISA effects are in reasonable agreement with the theoretical simulation of ISA. These experimental evidences make us believe that ISA resulting from the refractive index mismatch is most responsible for the self-formation of the void array. Obviously, it is promising to control the formation of the self-assembled void array by adjusting the amount of the spherical aberration. This is very attractive for precision machining of micro-devices that uses void as building blocks. The specific experiment is under way.

**Acknowledgement**

The authors greatly appreciate the experimental contribution from Fangfang Luo, Mengdi Qian, Junyi Ye. Thanks for the financial support from the National Natural Science Foundation of China (Grant Nos. 61205128, 11102075 and 51132004 ) and the Research Foundation for Advanced Talents of Jiangsu University (No. 09JDG022).